# Renormalization group flow and fixed point of the lattice topological charge in the 2–d O(3) $\sigma$–model[*]

Massimo D'Elia,

*Dipartimento di Fisica dell'Università and I.N.F.N.*

*Piazza Torricelli 2, I–56126 Pisa, Italy.*

Federico Farchioni and Alessandro Papa

*Institut für Theoretische Physik, Universität Bern*

*Sidlerstrasse 5, CH–3012 Bern, Switzerland.*

(November 1995, revised September 1996)

## Abstract

We study the renormalization group evolution up to the fixed point of the lattice topological susceptibility in the 2–d O(3) non–linear $\sigma$–model. We start with a discretization of the continuum topological charge by a local charge density, polynomial in the lattice fields. Among the different choices we propose also a Symanzik–improved lattice topological charge. We check step by step in the renormalization group iteration the progressive dumping of quantum fluctuations, which are responsible for the additive and multiplicative renormalizations of the lattice topological susceptibility with respect to the continuum definition. We find that already after three iterations these renormalizations are negligible and an excellent approximation of the fixed point is achieved. We also check by an explicit calculation that the assumption of slowly varying fields in iterating the renormalization group does not lead to a good approximation of the fixed point charge operator.

---

[*]Work supported in part by Fondazione "A. Della Riccia" (Italy).

## I. INTRODUCTION

In the field theoretical approach the lattice is an ultraviolet regulator for continuum quantum field theories, with the lattice spacing acting as cutoff; physical observables are constructed on the lattice in terms of local operators having the correct classical continuum limit. Two classes of effects hamper the extraction of physics from the lattice in this approach. First, discretized quantities are connected with the continuum ones through non–trivial renormalizations [1]; these renormalizations should be accurately evaluated. Second, a slight cutoff dependence of lattice quantities, not subtracted by the renormalization procedure and vanishing in the $a \to 0$ limit, induces systematic errors (the well known cutoff effects) in the determination of continuum quantities. Both classes of effects are artifacts of discretization.

A very elegant and radical solution to these problems comes from renormalization group (RG) theory [2]: all lattice actions and operators corresponding to the renormalized trajectory (i.e. the asymptotic flow line in the space of couplings under repeated RG transformations) are perfect, in the sense that they are free from lattice artifacts. Using perfect actions and perfect operators in numerical simulations, information about the continuum can be directly extracted from the lattice, since renormalizations and cutoff effects are absent. The first step in this ambitious program consists in constructing, for asymptotically free theories, fixed point (FP) actions and operators, which are perfect in the classical limit $g \to 0$ (being $g$ the asymptotically free coupling). A formal argument of perturbative RG implies that FP actions are 1–loop (quantum) perfect [3,4].

A method for the determination of FP actions for asymptotically free theories has been proposed in a recent paper [5]. The procedure has been applied to the 2–d O(3) non–linear $\sigma$–model on the lattice and a parametrization of the FP action $\mathcal{A}_{\text{FP}}$ suitable for numerical simulations has been found. In Ref. [6] a numerical implementation of RG has led to the definition of a FP topological charge operator in the geometrical approach [7] for the same model.

In this paper we build a FP topological charge operator for the O(3) σ–model in the framework of the field theoretical method [1]; this operator is used to extract the topological susceptibility of the model through numerical simulations. We adopt the FP parametrized action proposed in Ref. [5].

The paper is organized as follows: in Section II we review the techniques used in the construction of FP actions and operators and make some comments about the locality of renormalized and FP density operators; in Section III we outline the analytical determination of the FP topological charge at lowest order in a slowly varying field approximation; in Section IV we describe the numerical procedure which implements the RG transformations; by iterating this procedure we construct the RG flow of topological charge operators starting from a field theoretical definition and eventually converging to the FP operator; in Section V we outline the prescriptions of the field theoretical method in the computation of the topological susceptibility; we determine numerically its multiplicative renormalization and perturbative tail, when the first three renormalized charge operators of the RG flow are used; in Section VI, we study the scaling of the topological susceptibility by implementing the RG formalism in the usual MC techniques; in Section VII, we draw some conclusions.

## II. THE FP TOPOLOGICAL CHARGE OPERATOR

In order to clarify the further exposition and fix the notation, we briefly review the RG techniques used in constructing FP actions and operators for asymptotically free theories.

Let us consider an asymptotically free lattice spin theory. We write its partition function:

$$Z = \int [d\phi]\, e^{-\beta \mathcal{A}[\phi]} \;, \tag{1}$$

where $\phi$ is a spin field living on a lattice with spacing $a$, and $\beta \mathcal{A}[\phi]$ is some lattice regularization of the continuum action. In the case of the O(3) non–linear σ–model $[d\phi]$ would be the O(3)–invariant measure

$$[d\phi] = \prod_n d^3\phi_n \delta(\phi_n^2 - 1) \tag{2}$$

and $\beta\mathcal{A}[\phi]$ a regularization of the continuum action

$$\beta\mathcal{A}_{\text{cont}} = \frac{\beta}{2}\int d^2x\, \partial_\mu\phi(x)\cdot\partial_\mu\phi(x)\ , \qquad \phi^2(x) = 1\ . \tag{3}$$

A RG transformation with scale factor $\lambda$ for the spin system can be defined in the following way:

$$e^{-\beta'\mathcal{A}'[\Phi]} = \int [d\phi]\, e^{-\beta(\mathcal{A}[\phi]+\mathcal{T}[\Phi,\phi])}\ ; \tag{4}$$

here $\Phi$ is the blocked spin living on the lattice with spacing $a' = \lambda a$, related to a local average of the original spin variables; $\mathcal{T}[\Phi,\phi]$ is the blocking kernel, normalized in order to keep the partition function invariant under the transformation. The transformation used in Ref. [5] for the 2–d O(3) $\sigma$–model is obtained by dividing the original square lattice into $2\times 2$ blocks and assigning to each of them a new block spin $\Phi_{n_B}$. This spin is constructed by averaging the four original spins of the block. In the $\beta \to \infty$ limit, Eq. (4) can be solved in the saddle point approximation[1]:

$$\mathcal{A}'[\Phi] = \min_{\{\phi\}}\{\mathcal{A}[\phi] + \mathcal{T}[\Phi,\phi]\}\ . \tag{5}$$

In this limit the blocking kernel $\mathcal{T}[\Phi,\phi]$ turns out to be [5]:

$$\mathcal{T}[\Phi,\phi] = -\kappa\sum_{n_B}[\,\Phi_{n_B}\sum_{n\in n_B}\phi_n - |\sum_{n\in n_B}\phi_n|\,]\ , \tag{6}$$

where $\{n : n \in n_B\}$ are the sites $n$ of the original lattice belonging to the block labeled with $n_B$; $\kappa$ is a free parameter. The fixed point $\mathcal{A}^*$ of Eq. (5) is the starting point, on the $\beta = \infty$ critical surface, of the renormalized trajectory, which defines lattice actions free of cutoff effects. As a consequence, the FP action $\mathcal{A}_{\text{FP}} = \beta\mathcal{A}^*[\phi]$ is perfect in the classical limit $\beta \to \infty$, i.e. it is the classical perfect action.

---

[1] $\beta' = \beta - O(1)$ owing to asymptotic freedom.

For a given configuration $\{\Phi\}$ the value of the renormalized action $\mathcal{A}'[\Phi]$ can be determined by finding the configuration $\{\phi\}$, defined on the finer lattice, which minimizes the right hand side of Eq. (5). Furthermore, if the RG transformation is iterated $k$ times, the equation corresponding to Eq. (5) is

$$\mathcal{A}^{(k)}[\Phi] = \min_{\{\phi^{(1)},\ldots,\phi^{(k)}\}} \{\mathcal{A}^{(0)}[\phi^{(k)}] + \mathcal{T}[\phi^{(k-1)}, \phi^{(k)}] + \ldots + \mathcal{T}[\Phi, \phi^{(1)}]\} \; , \tag{7}$$

where $\phi^{(k)}$ is the field at the $k$–th fine level; starting on the finest lattice with any regularization $\mathcal{A}^{(0)}$ of the lattice action, a good approximation of the FP action can be obtained even for small values of $k$ [5].

In the case of the O(3) $\sigma$–model the parametrization used in Ref. [5] for $\mathcal{A}_{\mathrm{FP}}$ is:

$$\mathcal{A}_{\mathrm{FP}}[\phi] = \beta \left\{ -\frac{1}{2} \sum_{n,r} \rho(r)(1 - \phi_n \cdot \phi_{n+r}) \right. \tag{8}$$
$$\left. + \sum_{n_1,n_2,n_3,n_4} c(n_1, n_2, n_3, n_4)(1 - \phi_{n_1} \cdot \phi_{n_2})(1 - \phi_{n_3} \cdot \phi_{n_4}) + \quad \ldots \right\} \; ,$$

where $\rho$ represents the perfect discretization of the Laplacian; the coefficients $c(n_1, n_2, n_3, n_4)$ and $\rho(r)$ can be determined analytically by expanding the FP equation for slowly varying fields [5]. In the following we will always refer to the 24–couplings parametrization given in Table 4 of Ref. [5], where the couplings above the quartic order in the fields have been determined numerically by a fit procedure on the functional $\mathcal{A}^*[\phi]$, implementing Eq. (7) for an ensemble of configurations $\{\Phi\}$.

The RG transformation for an operator $O[\phi]$ is defined as follows:

$$O'[\Phi] \, e^{-\beta' \mathcal{A}'[\Phi]} = \int [d\phi] \, O[\phi] \, e^{-\beta(\mathcal{A}[\phi] + \mathcal{T}[\Phi,\phi])} \; ; \tag{9}$$

in the $\beta \to \infty$ limit it is straightforward to derive:

$$O'[\Phi] \;=\; O[\, \phi_{\min}[\Phi] \,] \; , \tag{10}$$

where $\{\phi_{\min}[\Phi]\}$ is the solution of the saddle point equation (5). Eq. (10) says that the renormalized operator $O'$ calculated on the blocked configuration $\{\Phi\}$ has the same value as the original operator $O$ calculated on the minimizing configuration $\{\phi_{\min}[\Phi]\}$. The iteration of Eq. (10) yields a succession of operators $O^{(k)}$ starting from some lattice regularization $O^{(0)}$, satisfying the recursive equation:

$$O^{(k)}[\Phi] = O^{(k-1)}[\phi_{\min}[\Phi]] \ . \tag{11}$$

The solution of (11) is:

$$O^{(k)}[\Phi] = O^{(0)}[\phi^{(k)}[\Phi]] \ , \tag{12}$$

being $\{\phi^{(k)}[\Phi]\}$ the minimizing configuration on the finest lattice in Eq. (7). The fixed point operator $O_{\text{FP}}$ is the $k \to \infty$ limit of $O^{(k)}$. A numerical determination of this limit can be practically viable if the FP operator is approximated by the renormalized operator after few iterations. Using the same RG argument of the FP action, one concludes that the FP operator is the classical perfect operator and its classical properties are the same of the corresponding continuum operator.

In the following we will be concerned with the topological charge operator of the O(3) non–linear $\sigma$–model; however the discussion is quite general and can be extended to every operator which can be expressed in terms of a local density.

Let us suppose to be on the fixed point $\mathcal{A}^*$ of the RG transformation (5). The minimizing configuration $\{\phi_{\min}[\Phi]\}$ on the fine lattice can be parametrized [6] in the following way:

$$(\phi_{\min})_n = \sum_{n_{\text{B}}} \alpha(n, n_{\text{B}}) \, \Phi_{n_{\text{B}}} + \sum_{l_{\text{B}}, m_{\text{B}}, n_{\text{B}}} \beta(n, l_{\text{B}}, m_{\text{B}}, n_{\text{B}})(1 - \Phi_{m_{\text{B}}} \cdot \Phi_{n_{\text{B}}})\Phi_{l_{\text{B}}} + \ldots \ . \tag{13}$$

Since we sit on the fixed point, the form of the parametrization is independent of the RG step. We define $d(n, m_{\text{B}})$ as the distance in coarse lattice units between the site $m_{\text{B}}$ and the site $n_{\text{B}}$ to which $n$ belongs. We define the *spread* of the parametrization (13) as the maximum distance $d(n, m_{\text{B}})$, over all the $m_{\text{B}}$'s in the coarse lattice involved in the parametrization of $(\phi_{\min})_n$. A good parametrization of $(\phi_{\min})_n$ can be possible even with a little spread (in Ref. [5] it was found a parametrization with spread $\sqrt{2}$). We also define the spread of a discretized local density operator $O[\phi](n)$ as the maximum distance $d(n, n')$ over the sites $n'$ involved in the definition of $O[\phi](n)$. By definition of local operator, this spread must keep finite in lattice units as the continuum limit is reached.

Let us now consider Eq. (11) for the topological charge and suppose that $Q^{(k-1)}$ is expressed in terms of a local density operator:

$$Q^{(k)}[\Phi] = \sum_n Q^{(k-1)}[\phi_{\min}[\Phi]](n) \ . \tag{14}$$

We can rewrite the last equation performing a blocking on the last summation:

$$Q^{(k)}[\Phi] = \sum_{n_B} \left( \sum_{n \in n_B} Q^{(k-1)}[\phi_{\min}[\Phi]](n) \right) = \sum_{n_B} Q^{(k)}[\Phi](n_B). \tag{15}$$

The last equation defines implicitly the density $Q^{(k)}(n_B)$. Suppose that the field parametrization has spread $r$ (in units of the coarse lattice spacing) and that $Q^{(k-1)}(n)$ has spread $s_{k-1}$ (in terms of the finer lattice spacing). Then, it is quite clear that $Q^{(k)}(n_B)$ has spread $s_k \sim r + \frac{1}{2} s_{k-1}$. From this follows that, starting from a lattice regularization $Q^{(0)} = \sum_n Q^{(0)}(n)$ with spread $s$, every element of the succession $Q^{(k)}$ can be parametrized in terms of a discretized local density having spread

$$s_k \sim r + \frac{r}{2} + \ldots + \frac{r}{2^{k-1}} + \frac{s}{2^k} \ . \tag{16}$$

Since $\phi_{\min}[\Phi]$ can be parametrized in a local way [6], the operators $Q^{(k)}$, $k = 1, 2, \ldots$, and ultimately the FP charge $Q_{\rm FP}$ itself, will still be expressed in terms of local densities $Q^{(k)}(n)$ with *finite* spread, the limiting value being $s_{\rm FP} \sim 2r$. This feature allows the application of heating techniques to determine the behavior of the renormalizations of the lattice operators $Q^{(k)}$ (see the following). What however is not under control is the number of powers of the fields contained in the density operators. Indeed, due to the non–linear terms in the field parametrization (13), the iteration of Eq. (11) is likely to give rise to an uncontrolled growth of the powers of the fields involved in the parametrization of $Q^{(k)}(n)$ as $k$ grows. That $Q_{\rm FP}$ in particular cannot be well approximated by a polynomial expression in the fields, can be concluded using the following two arguments.

i) The FP charge operator has, as in the continuum, a discrete spectrum. This is clear considering Eq. (12) in the case $O \equiv Q$: in the limit $k \to \infty$ the minimizing configuration on the finest lattice $\phi^{(k)}[\Phi]$ approaches a classical continuum solution [6], on which every starting regularization of the charge $Q^{(0)}$ gives an integer value (the same for all). This property is consistent with the RG argument which ensures that $Q_{\rm FP}$ is a perfect classical charge operator: the FP charge reflects, up to a minimum size, the discrete classical spectrum of the continuum operator

$$Q = \int d^2x \, Q(x) \, , \quad Q(x) = \frac{1}{8\pi} \epsilon_{\mu\nu} \, \epsilon_{ijk} \phi_i(x) \partial_\mu \phi_j(x) \partial_\nu \phi_k(x) \quad . \tag{17}$$

ii) On the lattice, differently from the continuum, each configuration can be continuously deformed into any other[2]. Joining arguments i) and ii) one concludes that $Q_{\rm FP}$ cannot have an analytical functional dependence in the lattice field[3]: one can produce for instance a discontinuous jump of $Q_{\rm FP}$ by continuously deforming a $Q_{\rm FP} = 0$ configuration into a $Q_{\rm FP} = 1$ one.

### III. $Q_{\rm FP}$ FOR SLOWLY VARYING FIELDS

A possible way to construct FP actions and operators, is to solve Eq. (11) making an expansion for slowly varying fields. The parameter of this expansion is the mean angle between the spins in the configuration $\{\phi\}$. In this Section we present the analytical determination of $Q_{\rm FP}[\phi]$ at lowest order. A necessary step in the calculation is finding the minimizing configuration $\{\phi_{\min}[\Phi]\}$ at lowest order in the expansion, i.e. the term linear in $\Phi$ in the parametrization (13). This amounts to solving the saddle point equation (5) in the free field theory case [8,5]. We express $\alpha(n, n_{\rm B})$ of Eq. (13) in Fourier transform:

$$\alpha(n, n_{\rm B}) = \int_{\rm B} \frac{d^2k}{(2\pi)^2} \, F(k) \, e^{i(n-n_{\rm B}) \cdot k} \quad , \tag{18}$$

where the subscript B means that the integration domain is the first Brillouin zone $[-\pi/a, \pi/a] \times [-\pi/a, \pi/a]$ and both $n$ and $n_{\rm B}$ are expressed in the coordinates of the coarse

---

[2]On the continuum this is impossible for starting and final configurations having different topological charge.

[3]In this respect, $Q_{\rm FP}$ resembles the geometrical definition of the lattice topological charge [7].

lattice. The result can be written as follows:

$$F(k) = \frac{1}{4} \alpha(k) \, \alpha'(2k)^{-1} \prod_{\mu=1}^{2} 2\cos\frac{q_\mu}{2} \; ; \tag{19}$$

here, $\alpha(k)$ is the inverse of the Laplacian $\rho(k)$ and the prime means after one RG transformation.

We used the following O(3)–invariant odd–parity parametrization of the topological charge operator at lowest order in the fields:

$$\begin{aligned}
8\pi Q &= \sum_{n_1,n_2,n_3} \mu(n_1 - n_2, n_1 - n_3) \, \epsilon_{ijk} \, \phi_i(n_1) \, \phi_j(n_2) \, \phi_k(n_3) \tag{20} \\
&= \int_B \frac{d^2 q_1}{(2\pi)^2} \frac{d^2 q_2}{(2\pi)^2} \, \mu(q_1, q_2) \, \epsilon_{ijk} \, \phi_i(-q_1 - q_2) \, \phi_j(q_1) \, \phi_k(q_2) \\
&= \left(\frac{1}{2^2}\right)^2 \sum_{m_1,m_2} \int_{B'} \frac{d^2 q_1}{\pi^2} \frac{d^2 q_2}{\pi^2} \, \mu(q_1 + m_1\pi, q_2 + m_2\pi) \\
&\quad \times \epsilon_{ijk} \, \phi_i(-q_1 - m_1\pi - q_2 - m_2\pi) \, \phi_j(q_1 + m_1\pi) \, \phi_k(q_2 + m_2\pi) \; ,
\end{aligned}$$

where $m_1, m_2$ are 2–vectors whose components take the values 0 and 1; the second equation follows from expressing the summation over the sites $n$ as a summation over the blocked sites $n_B$ and over the four sites of the block defining the fixed $n_B$; B′ means the "halved" Brillouin zone $[-\pi/(2a), \pi/(2a)] \times [-\pi/(2a), \pi/(2a)]$. If we now write $Q$ after one RG transformation as

$$8\pi Q' = \int_{B'} \frac{d^2 q_1}{\pi^2} \frac{d^2 q_2}{\pi^2} \, \mu'(q_1, q_2) \, \epsilon_{ijk} \, \Phi_i(-q_1 - q_2) \, \Phi_j(q_1) \, \Phi_k(q_2) \tag{21}$$

and, according to Eq. (14), plug the lowest order approximation of $\phi_{\min}[\Phi]$ (in Fourier transform) in the last integral in Eq. (20), we find:

$$\begin{aligned}
\mu'(2q_1, 2q_2) &= \frac{1}{2^2 2^2 4^3} \sum_{m_1,m_2} \mu(q_1 + m_1\pi, q_2 + m_2\pi) \tag{22} \\
&\quad \times \frac{\rho'(2q_1 + 2q_2) \rho'(2q_1) \rho'(2q_2)}{\rho(q_1 + m_1\pi + q_2 + m_2\pi) \rho(q_1 + m_1\pi) \rho(q_2 + m_2\pi)} \\
&\quad \times \prod_\mu 2\cos\left(\frac{q_1 + m_1\pi + q_2 + m_2\pi}{2}\right)_\mu \prod_\nu 2\cos\left(\frac{q_1 + m_1\pi}{2}\right)_\nu \prod_\sigma 2\cos\left(\frac{q_2 + m_2\pi}{2}\right)_\sigma .
\end{aligned}$$

The FP $\mu$–function, $\mu_{\mathrm{FP}}$, can be found iterating the previous relation; it comes out that

$$\mu_{\mathrm{FP}}(q_1, q_2) = \sum_{m_1, m_2 \in Z^2} \mu_{\mathrm{cont}}(q_1 + 2m_1\pi, q_2 + 2m_2\pi) \qquad (23)$$

$$\times \frac{\rho_{\mathrm{FP}}(q_1 + q_2)\rho_{\mathrm{FP}}(q_1)\rho_{\mathrm{FP}}(q_2)}{(q_1 + 2m_1\pi + q_2 + 2m_2\pi)^2(q_1 + 2m_1\pi)^2(q_2 + 2m_2\pi)^2}$$

$$\times \prod_\mu \frac{\sin((q_1 + q_2)/2)_\mu}{((q_1 + q_2)/2 + m_1\pi + m_2\pi)_\mu} \prod_\nu \frac{\sin(q_1/2)_\nu}{(q_1/2 + m_1\pi)_\nu} \prod_\sigma \frac{\sin(q_2/2)_\sigma}{(q_2/2 + m_2\pi)_\sigma} \;,$$

where $\mu_{\mathrm{cont}}(p, q) = \epsilon_{\mu\nu} p_\mu q_\nu$ and, using $\rho_{\mathrm{FP}}$, we have assumed that the action is also at the FP. We give for completeness the expression of $\rho_{\mathrm{FP}}$ [8,5]:

$$\rho_{\mathrm{FP}}^{-1} = \sum_{l \in Z^2} \frac{1}{(p + 2l\pi)^2} \prod_\mu \frac{\sin^2(p_\mu/2)}{(p_\mu/2 + l_\mu)^2} + \frac{1}{3\kappa} \;. \qquad (24)$$

As in [5] we use $\kappa = 2$. In Table I we show the coefficients $\mu(m, n)$ up to $O(10^{-5})$ (coefficients which can be related to these by symmetry transformations are omitted). The property of locality of $\mu(m, n)$ is consistent with the general discussion of previous Section about the spread of FP operators.

We have tested the leading $Q_{\mathrm{FP}}$ operator in the slowly varying field expansion, observing that it fails to reproduce the integer–valued spectrum of the continuum even on quite smooth configurations. We conclude that even on these configurations the neglected higher order operators dominate the expansion. This agrees with the theoretical conclusion at the end of the previous Section.

## IV. NUMERICAL DETERMINATION OF $Q_{\mathrm{FP}}$

An alternative way to face the problem of the determination of the FP charge of a given lattice configuration, is to solve the FP equation for the charge operator numerically. This approach has been described and applied recently to the O(3) $\sigma$–model in Ref. [6] for the geometrical definition of the lattice topological charge. Starting from a certain configuration $\{\Phi\}$ defined on the "coarse lattice" $L^{(0)}$, the minimizing configuration $\{\phi^{(k)}[\Phi]\}$ of Eq. (12) is singled out by constructing the succession of configurations defined on finer and finer lattices:

$$\{\Phi\}, \{\phi^{(1)}\}, \ldots, \{\phi^{(k)}\} \equiv \{\phi^{(k)}[\Phi]\} \;; \qquad (25)$$

given the configuration $\{\phi^{(k-1)}\}$ on the lattice $L^{(k-1)}$, the configuration $\{\phi^{(k)}\}$ on the lattice $L^{(k)}$ is determined by numerically minimizing the quantity

$$\mathcal{A}^*[\phi'] + \mathcal{T}[\phi^{(k-1)}, \phi'] \tag{26}$$

over all the configurations $\{\phi'\}$ living on the lattice $L^{(k)}$. After few steps of this iteration, the irrelevant content in $Q^{(k)}[\Phi] = Q^{(0)}[\,\phi^{(k)}[\Phi]\,]$ ($Q^{(0)}$ is the starting lattice operator) is negligible, and this quantity is a good approximation of $Q_{\rm FP}[\Phi]$. We have performed this program using three different starting lattice regularization $Q^{(0)}$ of the charge:

(1) the naïve regularization $Q_{\rm st}^{(0)}$:

$$Q_{\rm st}^{(0)}(x) \;=\; \frac{1}{8\pi}\, \epsilon_{\mu\nu}\, \epsilon_{ijk}\, \phi_i(x) \nabla_\mu \phi_j(x) \nabla_\nu \phi_k(x) \;, \tag{27}$$

where $\nabla_\mu$ is the symmetrized lattice derivative

$$\nabla_\mu \phi_i(x) \;=\; \frac{1}{2}(\phi_i(x+\mu) - \phi_i(x-\mu)) \;; \tag{28}$$

(2) the Symanzik tree-level improved (up to $O(a^6)$) operator $Q_{\rm Sym}^{(0)}$:

$$Q_{\rm Sym}^{(0)}(x) \;=\; \frac{1}{8\pi}\, \epsilon_{\mu\nu}\, \epsilon_{ijk}\, \phi_i(x) D_\mu \phi_j(x) D_\nu \phi_k(x) \;, \tag{29}$$

where $D_\mu$ is the improved lattice derivative:

$$D_\mu = \frac{1225}{1024} \nabla_\mu^{(1)} - \frac{245}{1024} \nabla_\mu^{(3)} + \frac{49}{1024} \nabla_\mu^{(5)} - \frac{5}{1024} \nabla_\mu^{(7)} \;, \tag{30}$$

$$\nabla_\mu^{(n)} \phi_i(x) \;=\; \frac{1}{2n}(\phi_i(x+n\mu) - \phi_i(x-n\mu)) \;;$$

(3) the FP operator at lowest order in the slowly varying field approximation $Q_{\rm SVF}^{(0)}$ derived in the previous Section[4].

We performed the minimization of the quantity in Eq. (26) by a local Metropolis algorithm accepting only lowering changes. The algorithm started from the minimizing configuration at lowest (linear) order in the slowly varying fields expansion:

---

[4]Using the first 12 leading couplings.

$$(\phi_{\text{start}}^{(k)})_n = \sum_{n_B} \alpha(n, n_B) \, \phi_{n_B}^{(k-1)} \ . \tag{31}$$

The number of minimization sweeps needed to reach the minimum decreases very rapidly with $k$ and at the third step no further minimization beyond the linear approximation is required at all: indeed, the configuration $\{\phi^{(2)}\}$ is generally very smooth and the lowest order approximation (31) allows to determine $\{\phi^{(3)}\}$ with sufficient accuracy.

The starting topological charges give clearly different results at the step zero, since large cutoff effects are present which are unequally handled by the various definitions; but, under iteration of the RG transformation, these UV effects are gradually erased and the three different field theoretical definitions of the lattice charge operator converge to the same operator $Q_{\text{FP}}$.

## V. THE FIELD THEORETICAL METHOD

The problem of extracting from the lattice the topological susceptibility of the O(3) $\sigma$–model has been studied for many years [6,7,9–15] using different approaches. The topological susceptibility has the dimension of a mass squared, so *admitting* that it is a physical quantity, i.e. renormalization group invariant, its (adimensional) lattice regularization should scale in the continuum limit with $\xi^{-2}$, where $\xi$ is the correlation length in lattice units. In fact, this scenario is obscured by a prevision of the semiclassical approximation [16] and some numerical evidences [17,18] which indicate that the topology of the model is UV dominated.

In the field theoretical method [1] a lattice topological charge density operator is defined as a local operator $Q(x)$ having the appropriate classical continuum limit. The matrix elements of $Q(x)$ are related to those in the continuum through a finite multiplicative renormalization:

$$Q(x) = a^2 \, Z(\beta) \, Q_{\text{cont}}(x) + O(a^4) \ . \tag{32}$$

The lattice version of the topological susceptibility is simply given by $\chi = 1/L^2 \, \langle (\, \sum_x Q(x) \,)^2 \rangle$ ($L$ is the lattice size). It is connected to the continuum counterpart by the relation

$$\chi(\beta) = a^2 Z(\beta)^2 \chi_{\text{cont}} + P(\beta) \langle I \rangle + O(a^4) ; \tag{33}$$

where $P(\beta)$ is the perturbative expectation value of the lattice topological susceptibility (the so called perturbative tail)[5]. Both $Z(\beta)$ and $P(\beta)$ are artifacts of discretization induced by the lattice quantum fluctuations. Eq. (33) allows to extract $\chi_{\text{cont}}$, the topological susceptibility of the continuum theory, from MC determinations of the lattice regularized susceptibility $\chi$. An intermediate step is the evaluation of the renormalizations $Z(\beta)$ and $P(\beta)$. In standard perturbation theory they are calculated as power series in $1/\beta$ [13]; an alternative approach is the Monte Carlo "heating" technique [19,13,18] (see the following).

Let's now come to the central issue of this paper, i.e. the application of the field theoretical method in combination with the FP lattice regularization of the topological charge. We move towards the FP operator through the operators $Q^{(k)}$ ($k = 1, 2, 3$) and construct the lattice susceptibilities $\chi^{(k)} = 1/L^2 \left\langle \left( \sum_x Q^{(k)}(x) \right)^2 \right\rangle$. The operators $Q^{(k)}$ still satisfy (as the starting operator $Q^{(0)}$) the prescriptions of field theory, since they can be expressed in terms of a local charge density with finite spread (see Section II). This allows us to write:

$$\chi^{(k)}(\beta) = a^2 Z^{(k)}(\beta)^2 \chi_{\text{cont}} + P^{(k)}(\beta) \langle I \rangle + O(a^4) . \tag{34}$$

The evaluation in perturbation theory of the renormalizations $Z^{(k)}$ and $P^{(k)}$ is not practicable in this case because the analytical dependence of the operators $Q^{(k)}$ on the field $\phi$ is not known; consequently, we turn to the numerical approach: the heating method.

The heating method [20,19] consists (for details see also [13,18]) in constructing on the lattice ensembles of configurations $\{C_t\}$, each configuration of the ensemble being obtained by performing a sequence of $t$ local Monte Carlo sweeps starting from a discretized classical configuration $C_0$ – a large instanton or a flat configuration. For small $t$, the heating process thermalizes only the small–range fluctuations which are responsible for the renormalizations:

---

[5]In the r.h.s. of Eq. (33) the mixing of the topological susceptibility with the action density has not been considered, since it is numerically negligible [18,15].

when the starting configuration is a large instanton (flat configuration), measuring $Q$ (or $\chi$) on the ensembles $\{C_t\}$ a plateau at the value of $Z(\beta)$ ($P(\beta)$) is observed after a certain time, not depending on $\beta$, corresponding to the thermalization of quantum fluctuations. The adaptation of this technique to the present case needs the building of the ensembles $\{C_t^{(k)}\}$ at the three fine levels. Starting from a configuration on the coarse lattice $\{\Phi\} \in \{C_t\}$, the corresponding configuration at the $k$–th fine level $\{\phi^{(k)}[\Phi]\} \in \{C_t^{(k)}\}$ was obtained by performing the minimization procedure described in Section IV. Averaging the starting definition of the topological charge (susceptibility) on the ensembles $\{C_t^{(k)}\}$ amounts to performing the heating procedure for $Q^{(k)}$ ($\chi^{(k)}$), so that the renormalization constant $Z^{(k)}$ ($P^{(k)}$) at each RG level can be determined. In Tables II – III we show the values of $Z^{(k)}$ and $P^{(k)}$ $k = 1, 2, 3$ (in the case $\beta = 1.15$ also $k = 0$) for five different $\beta$'s with the Symanzik improved topological charge as the starting operator: it is interesting to observe that $Z$ and $P$ converge rapidly towards 1 and 0 respectively when $k$ increases, thus indicating that the FP topological charge and susceptibility are insensitive to quantum fluctuations.

In Figs. 1 and 2 we show the behavior of $\langle Q^{(0)} \rangle (\beta = 1.15)$ and $\langle Q^{(1)} \rangle (\beta = 1.15)$ during the heating of an extended 1–instanton (with fixed boundary conditions). The most important observation is that the time of thermalization of the quantum fluctuations of $Q^{(1)}$ – not yet erased by the RG procedure – is very short ($\sim 4$ heating steps), thus indicating that the spread of this operator is small (of the order of few lattice spacings). This gives a numerical support to the arguments of Section II about the spread of renormalized operators defined starting from a local density.

From the analysis of the data of heating we conclude that $Q_{\rm Sym}^{(0)}$ is the starting operator closest to the fixed point, since it yields at a given $k$ the values of $Z$ and $P$ closest to 1 and 0 respectively (see Table IV for $Z(\beta = 1.15)$) [6]. In the following we will be concerned about the data coming from this starting operator.

---

[6]That $Q_{\rm Sym}$ is closest to the FP operator could support the idea that Symanzik's improvement program goes in the same direction of the RG improvement.

## VI. MC SIMULATIONS

We performed Monte Carlo simulations on a starting "coarse" lattice $90 \times 90$ for five values of $\beta$ (1., 1.05, 1.10, 1.15, 1.20), $\xi$ ranging from 12 to 35 lattice units. We measured $Q_{\text{Sym}}^{(0)}$ (only in the case $\beta = 1.15$), $Q_{\text{Sym}}^{(1)}$, $Q_{\text{Sym}}^{(2)}$ and $Q_{\text{Sym}}^{(3)}$ using the numerical technique discussed in Section IV. We considered $\sim 5000$ decorrelated starting configurations at thermal equilibrium for each value of $\beta$.

In Table V and in Fig. 3 we present values of $\chi_{\text{cont}} \cdot \xi_p^2$, where $\xi_p \equiv \xi a$, at the five $\beta$'s considered on each RG level: $\chi_{\text{cont}}$ has been extracted according to Eq. (34); $\xi$ was determined by evaluating the exponential decay of the wall–wall correlation function on lattices satisfying the condition $L/\xi \simeq 7$ (in order to keep safe from finite size effects)[7].

Data in Table V allows two observations: the vertical reading shows that there is no scaling up to correlation length $\xi \simeq 35$; the horizontal reading shows that the physical signal $\chi_{\text{cont}}$ increases after each RG step. The first observation is a consequence of the UV dominated nature of the topology of the model, already well established in literature; the second effect is subtler, related to the different scale invariance properties of the three operators $Q^{(k)}$, $k = 1, 2, 3$. Being classically perfect, $Q_{\text{FP}}$ should have a scale invariant spectrum, so it should attribute the correct topological charge to lattice configurations up to a very small size, close to the lattice unit. In the case of $Q^{(k)}$ this property is only approximate, improving for increasing $k$. This is visually clear in Fig. 4, where we show the behavior of topological charge at the coarse, the first fine and the second fine levels on small size classical configurations (1–instantons with fixed boundary conditions). We can see that the topological charge at the coarse level underestimates the correct continuum value $Q = 1$ on classical configurations with small size, while the correct value is asymptotically attained as $k$ increases. We conclude that the slight dependence of our determination of $\chi_{\text{cont}}$ on the "fineness level" of the lattice, $k$, is an effect of the progressive saturation of the topological signal coming from small size instantons. We observe that at $k = 3$ this saturation can be considered practically complete; we argue therefore that only a negligible improvement would come from further RG steps. It is also evident from Fig. 4 that no topological signal is observable on the lattice below 0.9 lattice units, even when using a FP operator and a FP action. Since most of the topological signal in this model comes from the region of small lengths, a cutoff effect is still present in the lattice theory, which manifests itself in the unphysical dependence of $\chi_{\text{cont}}$ on $\xi$ (non–scaling).

---

[7]We are indebted to the authors of Ref. [6] for allowing us to use their cluster algorithm program for measurements of correlation length.

## VII. SUMMARY AND CONCLUSIONS

The theory of Wilson's renormalization group states the existence of perfect lattice operators, i.e. operators discretized on the lattice which are free of cutoff effects. These operators correspond to the renormalized trajectory of a given renormalization group transformation. The fixed point of the transformation defines the fixed point operator, the classically perfect lattice operator.

In this paper we have studied the evolution towards the fixed point of the two–point correlation function of the topological charge, i.e. the topological susceptibility, in the 2–d O(3) non–linear $\sigma$–model on the lattice. For this purpose, we have discretized the continuum topological charge on the lattice in terms of a local operator, polynomial in the lattice fields, and we have implemented the renormalization group procedure in a numerical fashion.

We have observed that the quantum fluctuations are progressively erased as the fixed point is approached, thus leading to an integer–valued spectrum of the topological charge on the lattice. The amount of quantum fluctuations is quantified by the renormalization constants $Z(\beta)$ and $P(\beta)$ entering the lattice definition of the topological susceptibility. We have measured these constants by the heating method observing their rapid disappearance with the iteration of the renormalization group transformation. The operators obtained at each step of the renormalization group procedure can be considered as improvements of the starting operators, in the sense of a suppression of the quantum noise around the physical signal. A similar improvement, consisting in the reduction of the renormalization constants, was obtained in Ref. [21] by using smearing techniques. We have also calculated the physical value of the topological susceptibility at thermal equilibrium by subtracting the effect of the renormalizations from the lattice signal at each renormalization group step. The physical quantity obtained increases with the renormalization group step up to convergence within a few percent of accuracy. This convergence corresponds to the saturation of the topological signal by smaller and smaller instantons, down to the minimum size allowed on the lattice. The physical signal at the fixed point does not exhibit a scaling behavior, since the topological contribution of instantons with size lower than the critical size, which is still lost on the lattice, is dominating.

A procedure similar to the one described in this paper has already been applied in Ref. [6] using the geometrical approach to the lattice topological charge: we observe that the two different methods, while diverging at the starting point, give asymptotically consistent results when the renormalization group procedure is carried on.

We have also presented an analytic determination of the fixed point topological charge in the approximation of slowly varying fields, by iterating the renormalization group transformation at the leading order. The resulting operator fails to reproduce the integer–valued spectrum of the continuum even on quite smooth configurations, thus leading to the conclusion that the neglected higher order terms are important. The negative result obtained is not completely unexpected since an analytic expression in the lattice fields for the topological charge operator cannot reproduce a discrete (and, therefore, discontinuous) spectrum on the lattice, where each configuration can be continuously transformed into any other.

The present paper can be considered as a preliminary work in view of the determination of the fixed point topological charge for the SU(3) gauge theory. For this theory, however, the numerical procedure applied in the present paper is no longer viable, owing to the huge computation time needed for the determination of the configurations on the finer lattices. We retain that a promising approach is to search for a (few–couplings) parametrization of the fixed point topological charge operator which works even on coarse lattices.

## ACKNOWLEDGMENTS

We wish to thank M. Blatter, R. Burkhalter, A. Di Giacomo, A. Hasenfratz, P. Hasenfratz, F. Niedermayer, for many useful discussions. This work was partially supported by Fondazione "A. Della Riccia" (Italy).

FIGURES

FIG. 1. Topological charge at the coarse level during the heating of an extended 1–instanton (with fixed boundary conditions) at $\beta = 1.15$. The line represent the fit (with errors) of the data on the plateau.

FIG. 2. Topological charge at the first fine level during the heating of an extended 1–instanton (with fixed boundary conditions) at $\beta = 1.15$. The line represent the fit (with errors) of the data on the plateau.

FIG. 3. $\chi \cdot \xi_p^2$ ($\xi_p \equiv \xi a$) for five correlation lengths $\xi$ at the various RG steps. The starting topological charge regularization is the Symanzik charge. Data at the coarse, 2nd and 3rd fine level have been slightly shifted for the sake of readability.

FIG. 4. Topological charge at the coarse, first fine and second fine levels on classical configurations consisting of 1–instantons (with fixed boundary conditions) with different size. Instantons have been smoothed out by some cooling steps before the charge measurement.

TABLES

TABLE I. Largest non–equivalent values of $\mu(m,n)$.

| $m$ | $n$ | $\mu(m,n)$ | $m$ | $n$ | $\mu(m,n)$ |
|---|---|---|---|---|---|
| (1,1) | (1,0) | $5.00100 \times 10^{-2}$ | (3,0) | (2,1) | $-4.35376 \times 10^{-5}$ |
| (2,0) | (1,1) | $-6.27316 \times 10^{-3}$ | (3,1) | (2,0) | $4.15550 \times 10^{-5}$ |
| (2,1) | (1,1) | $-4.90036 \times 10^{-3}$ | (3,2) | (2,2) | $3.90132 \times 10^{-5}$ |
| (2,1) | (2,0) | $1.52221 \times 10^{-3}$ | (3,2) | (2,1) | $-2.53691 \times 10^{-5}$ |
| (2,1) | (1,2) | $-9.39105 \times 10^{-4}$ | (3,1) | (2,2) | $2.20664 \times 10^{-5}$ |
| (2,1) | (2,–1) | $3.55142 \times 10^{-4}$ | (2,2) | (2,1) | $1.51901 \times 10^{-5}$ |
| (2,2) | (2,0) | $1.16270 \times 10^{-4}$ | (3,1) | (3,0) | $1.37893 \times 10^{-5}$ |
| (3,1) | (2,1) | $-8.39146 \times 10^{-5}$ | (3,2) | (3,1) | $-1.28112 \times 10^{-5}$ |

TABLE II. $Z(\beta)$ for the different $\beta$'s at the various RG levels, $k$. The starting operator is $Q_{\text{Sym}}^{(0)}$.

| $\beta$ | coarse | 1. level | 2. level | 3. level |
|---|---|---|---|---|
| 1.00 | | 0.892(9) | 0.980(5) | 0.9950(27) |
| 1.05 | | 0.910(8) | 0.990(3) | 0.9999(11) |
| 1.10 | | 0.921(6) | 0.9929(23) | 0.9992(8) |
| 1.15 | 0.450(6) | 0.929(6) | 0.9931(21) | 0.9990(6) |
| 1.20 | | 0.936(5) | 0.9939(15) | 0.9991(3) |

TABLE III. $P(\beta) \times 10^4$ for the different $\beta$'s at the various RG levels, $k$. The starting operator is $Q_{\text{Sym}}^{(0)}$.

| $\beta$ | coarse | 1. level | 2. level | 3. level |
| --- | --- | --- | --- | --- |
| 1.00 | | 0.571(22) | 0.177(7) | 0.0466(24) |
| 1.05 | | 0.179(10) | 0.0276(22) | 0.0037(5) |
| 1.10 | | 0.123(7) | 0.0241(19) | 0.0022(3) |
| 1.15 | 0.804(13) | 0.094(5) | 0.0116(10) | 0.00064(12) |
| 1.20 | | 0.066(3) | 0.0088(9) | 0.00061(11) |

TABLE IV. $Z(\beta)$ at $\beta = 1.15$ for the three different starting regularizations of the topological charge.

| | coarse | 1. level | 2. level | 3. level |
| --- | --- | --- | --- | --- |
| $Q_{\text{st}}^{(0)}$ | 0.411(4) | 0.855(5) | 0.9625(25) | 0.9903(10) |
| $Q_{\text{Sym}}^{(0)}$ | 0.450(6) | 0.929(6) | 0.9931(21) | 0.9990(6) |
| $Q_{\text{SVF}}^{(0)}$ | 0.384(3) | 0.815(5) | 0.9478(27) | 0.9860(11) |

TABLE V. Topological susceptibility in physical units up to the third RG level ($\xi_p \equiv \xi a$). For comparison, we also report in the last column the data of Ref. [6].

| $\beta$ | $\xi$ | $\chi_{\text{coarse}} \cdot \xi_p^2$ | $\chi_{\text{1. level}} \cdot \xi_p^2$ | $\chi_{\text{2. level}} \cdot \xi_p^2$ | $\chi_{\text{3. level}} \cdot \xi_p^2$ | $\chi_{\text{1. level}}^{\text{geom}} \cdot \xi_p^2$ [6] |
|---|---|---|---|---|---|---|
| 1.00 | 12.16(3) | | | | | 0.1448(15) |
| | 12.27(3) | | 0.137(7) | 0.150(6) | 0.151(5) | |
| 1.05 | 15.81(6) | | 0.161(8) | 0.172(6) | 0.173(5) | |
| 1.10 | 20.34(9) | | 0.173(8) | 0.185(7) | 0.188(6) | |
| | 20.40(9) | | | | | 0.1893(27) |
| 1.15 | 26.26(14) | 0.160(13) | 0.177(8) | 0.192(7) | 0.195(7) | |
| 1.20 | 34.13(25) | | 0.194(11) | 0.210(10) | 0.213(9) | |
| | 34.4(3) | | | | | 0.224(5) |

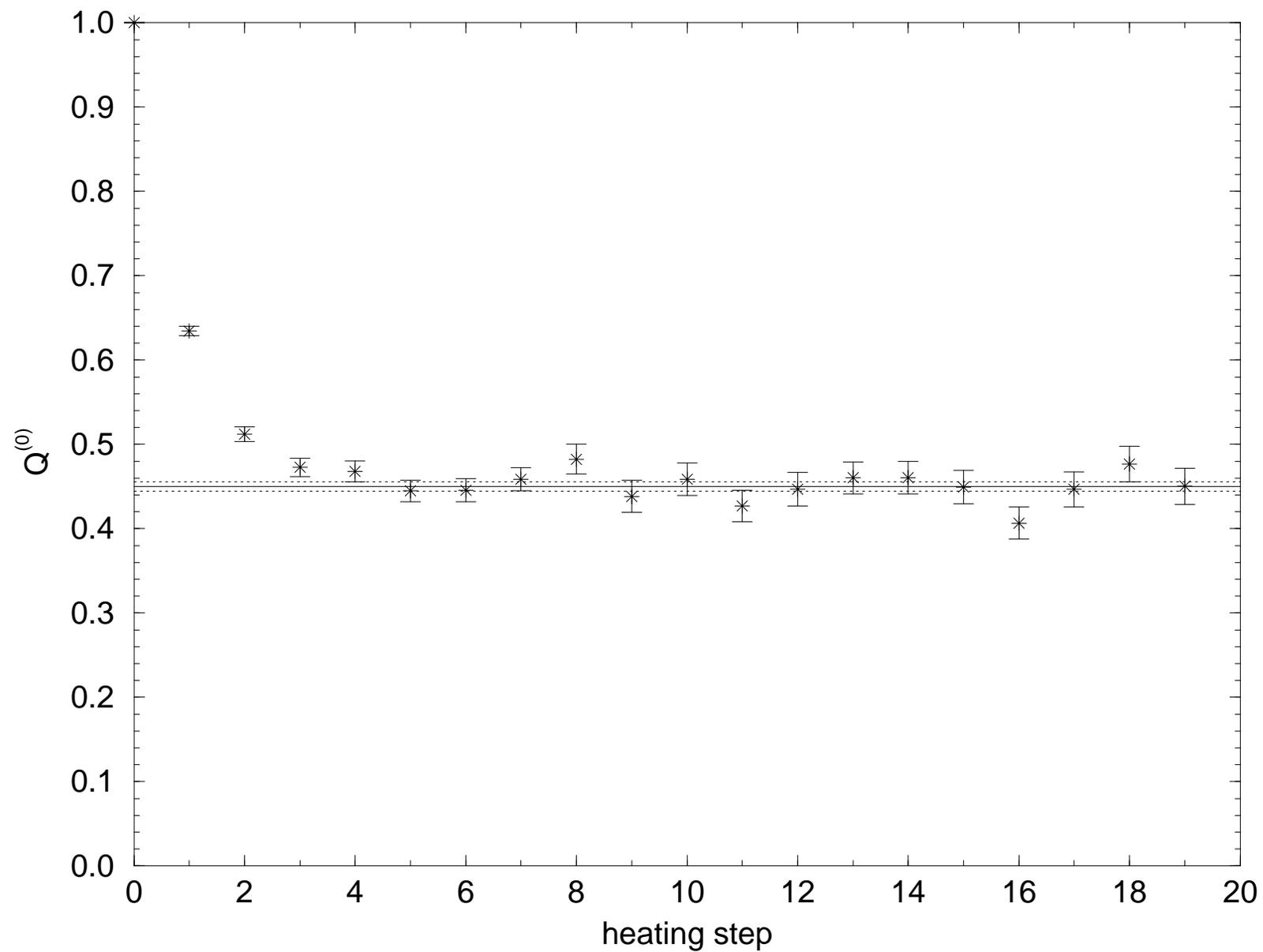

FIG. 1

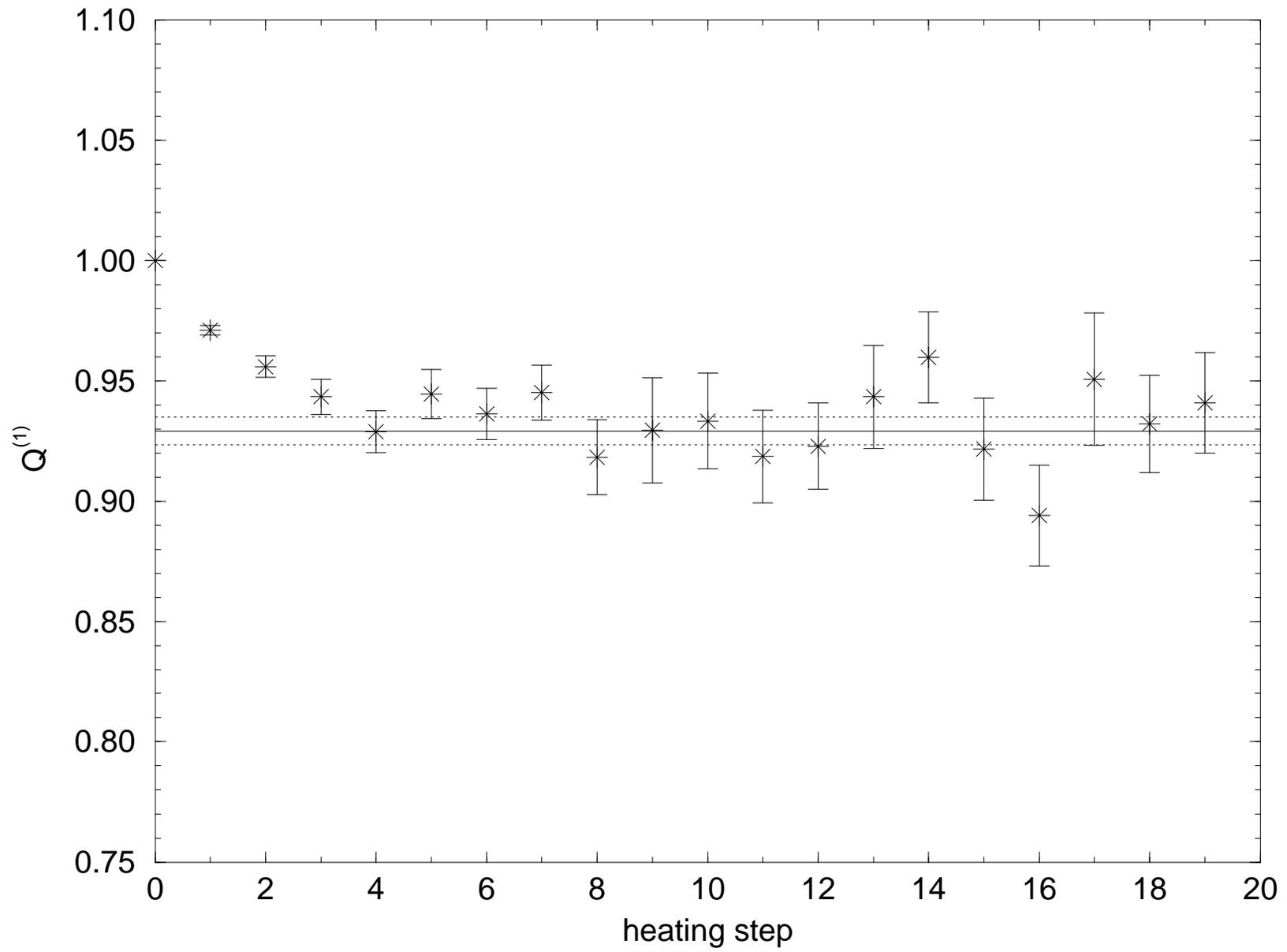

FIG. 2

FIG. 3

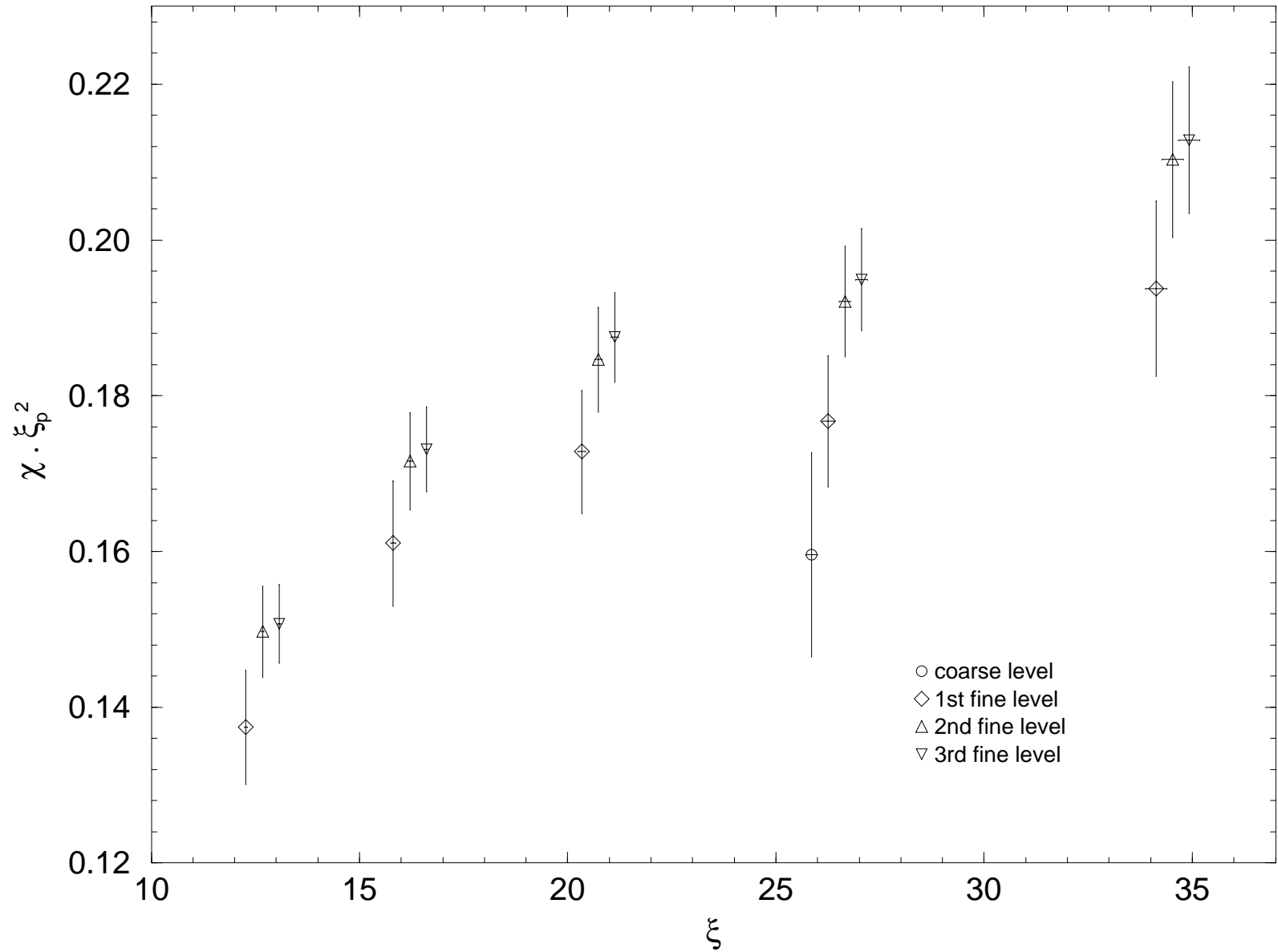

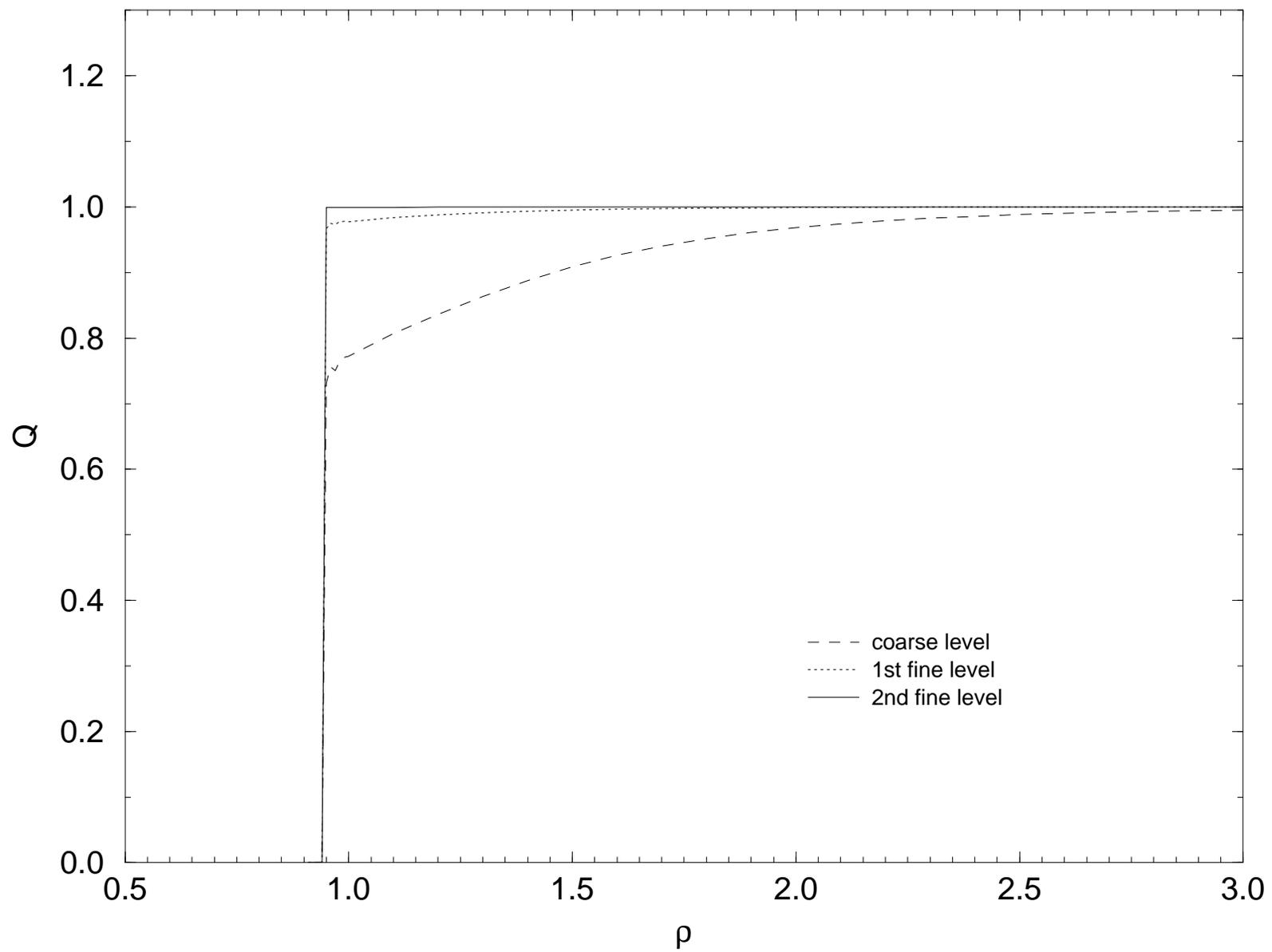
FIG. 4